\documentclass[twocolumn,english,prb,aps,superscriptaddress]{revtex4}
\usepackage{lmodern}

\usepackage[latin9]{inputenc}
\usepackage{amsmath}
\usepackage{amssymb}
\usepackage{graphicx}
\usepackage{esint}

\makeatletter

\usepackage[USenglish]{babel}

\DeclareMathOperator{\sgn}{sgn}
\newcommand{\beq}{\begin{equation}}
\newcommand{\eeq}{\end{equation}}
\newcommand{\bea}{\begin{eqnarray}}
\newcommand{\eea}{\end{eqnarray}}
\newcommand{\be}{\begin{equation}}
\newcommand{\ee}{\end{equation}}

\renewcommand{\phi}{\varphi}

\usepackage{babel}

\makeatother

\begin{document}

\title{Effect of pairing fluctuations on the spin resonance in Fe-based superconductors}

\author{Alberto Hinojosa}
\affiliation{Department of Physics, University of Wisconsin,
Madison, Wisconsin 53706, USA}
\author{Andrey V. Chubukov}
\affiliation{Department of Physics, University of Wisconsin,
Madison, Wisconsin 53706, USA}
\author{Peter W\"{o}lfle}
\affiliation{Institute for Condensed Matter Theory and Institute for Nanotechnology, Karlsruhe Institute of Technology, D-76128 Karlsruhe, Germany}

\begin{abstract}
The spin resonance observed in the inelastic neutron scattering data on Fe-based superconductors has played a prominent role in the quest for determining the symmetry of the order parameter in these compounds. Most theoretical studies of the resonance
   employ an RPA-type approach in the particle-hole channel and associate the resonance in the spin susceptibility $\chi_S (\mathbf{q}, \omega)$
   at momentum ${\bf Q} = (\pi,\pi)$ with the spin-exciton of an $s^{+-}$ superconductor, pulled below $2\Delta$ by residual attraction
  associated with the sign change of the gap between Fermi points connected by ${\bf Q}$.  Here we explore the effect of fluctuations in the particle-particle channel on the spin resonance.
     Particle-particle and particle-hole channels are coupled in a superconductor and to what extent the spin resonance can be viewed as a particle-hole exciton needs to be addressed. In the case of purely repulsive interactions we find that the particle-particle channel at total momentum ${\bf Q}$  (the $\pi $channel)
      contributes little to the resonance.
      However, if the interband  density-density interaction is attractive and the $\pi-$resonance is possible on its own,
       along with spin-exciton,  we find a much stronger
        shift of the resonance frequency from the position of the would-be spin-exciton resonance.  We also show that the expected double-peak structure
         in this situation does not appear because of the strong coupling between particle-hole and particle-particle channels, and $\mathrm{Im} \chi_S ({\bf Q}, \omega)$
           displays only a single peak.

\end{abstract}
\maketitle

\section{Introduction}

The spin resonance, observed by inelastic neutron scattering (INS)
experiments first in the cuprates\cite{rossat} and then in heavy-fermion~\cite{heavy_fermion} and Fe-based superconductors (FeSCs)\cite{claus}, has been the subject of intense theoretical and experimental studies over the past decade using both metallic~\cite{scalapino,norman,res_fe,morr} and near-localized strong-coupling scenarios~\cite{brinck} (see Ref. \cite{eschrig} for a review). The theoretical interpretations of the resonance can be broadly split into two classes. The first class of theories assumes that the spin resonance is a magnon, overdamped in the normal state due to the strong decay into low-energy particle-hole pairs, but emerging prominently in the superconducting state due to reduction of scattering at low energies~\cite{morr,gorkov}. In this line of reasoning the resonance energy $\Omega_{res}$ is the magnon energy and as such it is uncorrelated with the superconducting gap $\Delta$. However, the decay of magnons into particle-hole pairs is only suppressed at energies below $2\Delta$, hence the magnons become sharp in a superconductor only if their energy is below $2\Delta$.  The symmetry of the superconducting state does not play a crucial role here. It is only relevant that the superconducting gap is finite at the Fermi surface (FS) points connected by the gap momentum ${\bf Q}$.

Theories from the second class assume that the resonance does not exist in the normal state
 and emerges in the superconducting state as a feedback effect from superconductivity, like e.g., the Anderson-Bogolyubov mode in the case of a charge neutral single condensate
component, a Leggett mode in the case of several gap components,   a wave-like excitation of a spin-triplet order parameter, or a pair vibration mode in the case of a gap parameter possessing internal structure~\cite{peter_book}.
These ``feedback'' theories  can be further split into three sub-classes.  In the first the resonance is viewed as a spin-exciton, i.e., the pole in the
      dynamical spin susceptibility $\chi (\mathbf{q}, \Omega)$  dressed by multiple particle-hole bubbles~\cite{scalapino,norman,res_fe}. Such $\chi (\mathbf{q}, \Omega)$ can be obtained by using a computational scheme based on the random-phase approximation (RPA). It was argued that, if the superconducting gap changes sign between
 FS points connected by ${\bf Q}$, the residual attraction pulls the resonance frequency to $\Omega_{res} <2\Delta$, where the decay into particle-hole pairs is reduced below $T_c$ and vanishes at $T=0$. As a result, at $T=0$, $\chi^{''} (\mathbf{q}, \Omega)$ has a
  $\delta-$functional peak at $\Omega_{res}$.  In this respect, if the resonance is an exciton, its existence necessary implies that the superconducting gap changes sign either between patches of the FS connected by ${\bf Q}$  or between different Fermi pockets again connected by ${\bf Q}$.
   The role of the resonance in allowing to determine the structure of the gap in a number of different superconductors has been highlighted in \cite{scalapino12}.
   Theories of the second subclass explore the fact that in a superconductor the particle-hole and particle-particle channels are
 mixed and argue that the strongest resonance is in the particle-particle channel and the measurements of the spin susceptibility just reflect the ``leakage'' of this resonance into the particle-hole channel. The corresponding resonance has been labeled as the $\pi$-exciton \cite{zhang}, where the $\pi$ boson is a particle-particle
  excitation with total momentum ${\bf Q}$ (a ``pair density-wave'' in modern nomenclature~\cite{kiv_fra,lee,agterberg}).
    Finally, theories of the third class explore the possibility that the resonance emerges due to coupling between fluctuations in particle-hole and particle-particle channel. Within RPA, such resonance
   is due to non-diagonal terms in the generalized RPA which includes both particle-hole and particle-particle bubbles.  It was called a plasmon~\cite{macdonald}
    to stress the analogy with collective excitations in an electronic liquid.
		
 The interplay between the ``damped spin-wave'' scenarios, the spin-exciton $\pi-$resonance, and the plasmon scenarios for 2D high-$T_c$ cuprates has been studied in detail in the past decade. The outcome is that near and above optimal doping the resonance is best described as a spin-exciton,
  with relatively weak corrections from coupling to the particle-particle channel\cite{norman,fomin,hao}, while in the underdoped regime, where superconductivity emerges from a pre-existing pseudogap phase, both spin-exciton and spin-wave scenarios have been argued to account for the neutron resonance~\cite{eschrig}. The situation is more complicated in 3D heavy-fermion systems~\cite{gorkov,ilya_1} because there the excitonic resonance has a finite width even at $T=0$ if the locus of FS points separated by ${\bf Q}$  intersects  a line of gap nodes.

In this work we discuss the interplay between spin-exciton, $\pi-$resonance, and plasmon scenarios in FeSCs.
Previous studies of the resonance in FeSCs \cite{res_fe} focused only on the response in the particle-hole (spin-exciton) channel and neglected the coupling
 between particle-hole and particle-particle channels. Our goal is to analyze the effect of such coupling.

     As we said, the resonance peak has been observed below $T_c$ in several families of FeSCs\cite{claus}. The spin response above $T_c$ in FeSCs is rather featureless away from small doping, which implies that the magnetic excitations in the normal state are highly overdamped and don't behave as damped spin waves.
The  full analysis of the spin resonance in FeSCs is rather involved as these systems are multi-band materials with four or five FSs, on which the superconducting gap has different amplitudes and phases.  Still, the basic conditions for spin-excitonic resonance are the same as in the cuprates and heavy fermion materials: namely, the resonance emerges at momentum ${\bf Q}$ if the superconducting gap changes sign between FS points connected by ${\bf Q}$.  This condition holds if the superconducting gap has $s^{+-}$  symmetry, as most researchers believe, and changes sign between at least some hole and electron pockets. An alternative scenario~\cite{kontani}, which we will not discuss in this paper is that the superconducting state has a conventional, sign-preserving  $s^{++}$ symmetry, and the observed neutron peak is not a resonance but rather a hump at frequencies slightly above $2\Delta$.

In FeSCs that contain both hole and electron pockets, the
 resonance has been observed at momenta around ${\bf Q} = (\pi,\pi)$, which is roughly the distance between hole and electron pockets
  in the actual (folded) Brillouin zone.  To account for the resonance and, at the same time, avoid unnecessary complications, we consider
a minimal three-band model (one hole pocket and two electron pockets), and neglect the angular dependence of the interactions along the FSs.  Including this dependence and additional hole pockets will complicate the analysis but we don't expect it to lead to any qualitative changes to our results.

We find that for repulsive density-density and pair-hopping interactions, the resonance peak is, to a good accuracy, a spin-exciton.
  The $\pi-$resonance does not develop on its own, and the coupling between the resonant spin-exciton channel and the non-resonant $\pi-$channel only slightly shifts the
   energy of the excitonic resonance.  We also considered the case (less justified microscopically) when the interaction in the $\pi$ channel is attractive, such that
    both spin-exciton and $\pi$ resonance develop on their own at  frequencies below $2\Delta$. One could expect in this situation that the full dynamical spin susceptibility has two peaks. We found, however, that this happens only if we make the coupling between particle-hole and particle-particle channels
      artificially small. When we restored the original coupling, we  found, in general,  only one peak below $2\Delta$.  The peak is a mixture of a spin-exciton and $\pi$-resonance and at least in some range of system parameters its energy is smaller than that of a spin-exciton and a $\pi$-resonance. This implies that, when both channels are attractive, the coupling between the two plays a substantial role in determining the position of the true resonance which can, at least partly, be viewed as a plasmon. A somewhat similar result has been obtained earlier for the cuprates~\cite{norman,hao,macdonald}.
For some system parameters we
       did find two peaks in $\mathrm{Im} \chi_S ({\bf Q}, \omega)$, but for one of them $\mathrm{Im} \chi_S$
			 has wrong sign. We verified that this indicates that for such parameters the system is unstable either against condensation of $\pi$ excitations (i.e., against superconductivity at momentum ${\bf Q}$), or against the development of SDW order in co-existence with superconductivity.

The paper is organized as follows: In the next section we consider the model. In Sec. \ref{chi_rpa} we obtain the dynamical spin susceptibility within the generalized RPA scheme, which includes the coupling between particle-hole and particle-particle channels. In Sec. \ref{results} we analyze the profile of $\chi_S (\mathbf{Q}, \omega)$ first for purely repulsive density-density and pair-hopping interactions, and then for the case when we allow the density-density interaction to become attractive. We summarize our conclusions in Sec. \ref{concl}.

\section{The model}
The FeSCs are multiband metals with two or three hole FS pockets centered
around the $\Gamma$ point $(0,0)$ and two elliptical electron pockets
centered at $(\pi,\pi)$ in the folded BZ with two iron atoms per unit cell. The electron pockets are elliptical and related by symmetry, while the hole pockets are $C_4$-symmetric, but generally differ in size. Since we are only interested in studying the role of the particle-particle channel in the spin response function, for which the non-equivalence between hole pockets is not essential, we  consider the case of two hole pockets and assume that they are circular and identical, and also neglect the ellipticity of electron pockets. Under these assumptions our model reduces effectively to only one hole pocket ($c$
fermions) and one electron pocket ($f$ fermions). The fact that there are actually two hole and two electron pockets only adds up combinatoric factors.

 The free part of the
Hamiltonian is
\begin{equation}
H_0=\sum_{\mathbf{k},\sigma} \left( \xi^c_\mathbf{k} c^\dagger_{\mathbf{k}%
\sigma}c_{\mathbf{k}\sigma} +\xi^f_{\mathbf{k}+\mathbf{Q}} f^\dagger_{%
\mathbf{k}+\mathbf{Q}\sigma}f_{\mathbf{k}+\mathbf{Q}\sigma}\right),
\end{equation}
where
\begin{align}
\xi^c_\mathbf{k} &=\mu_c-\frac{k_x^2+k_y^2}{2m_c}, \\
\xi^f_{\mathbf{k}+\mathbf{Q}} &= -\mu_f +\frac{k_x^2 + k^2_y}{2m_f}
\end{align}

We do not study here how superconductivity develops from interactions, as that work has been done elsewhere~\cite{sc}.
 Instead, we simply
assume that the system reaches a superconducting state with $s^{+-}$ symmetry before it becomes unstable towards magnetism and
take the superconducting gaps as inputs. In this state, the free part of the Hamiltonian is
\begin{equation}
H_0^{SC}=\sum_{\mathbf{k},\sigma} \left( E^c_\mathbf{k} c^\dagger_{\mathbf{k}%
\sigma}c_{\mathbf{k}\sigma} +E^f_{\mathbf{k}+\mathbf{Q}} f^\dagger_{%
\mathbf{k}+\mathbf{Q}\sigma}f_{\mathbf{k}+\mathbf{Q}\sigma}\right),
\end{equation}
where the dispersions are $E^c_\mathbf{k}=\sqrt{(\xi^c_\mathbf{k}%
)^2+(\Delta^c)^2}$ and $E^f_{\mathbf{k}+\mathbf{Q}}=\sqrt{(\xi^f_{\mathbf{k}+\mathbf{Q}})^2+(\Delta^f)^2}$, and $\Delta^c=-\Delta^f\equiv \Delta$.

Now we consider interactions that contribute to the spin susceptibility. They consist of a density-density interband
interaction $u_1$ and a correlated interband hopping $u_3$.    Intraband repulsion
 only affects the chemical potentials
but does not otherwise contribute to the spin susceptibility. Interband exchange in principle contributes in the $\pi$ channel but in renormalization group analysis it flows to small values\cite{RG}.
In general, $u_1$ and $u_3$  depend on the angle in momentum space via coherence factors associated with the transformation from the orbital to the band basis~\cite{vavilov}.  However, this
complication is not essential for our purposes and we take both
interactions to be momentum independent. The interaction Hamiltonian is
\begin{align}
H_{\mathrm{int}}& =u_{1}\sum_{[1234],\sigma \neq \sigma ^{\prime
}}c_{\mathbf{p}_1\sigma }^{\dagger }f_{\mathbf{p}_2\sigma ^{\prime }}^{\dagger }f_{\mathbf{p}_3\sigma
^{\prime }}c_{\mathbf{p}_4\sigma } \nonumber\\
& +u_{3}\sum_{[1234],\sigma \neq \sigma ^{\prime }}\left( c_{\mathbf{p}_1\sigma
}^{\dagger }c_{\mathbf{p}_2\sigma ^{\prime }}^{\dagger }f_{\mathbf{p}_3\sigma ^{\prime
}}f_{\mathbf{p}_4\sigma }+\mathrm{h.c.}\right),
\end{align}%
where the sum over momenta obeys momentum conservation as usual ($\mathbf{p}_1+\mathbf{p}_2=\mathbf{p}_3+\mathbf{p}_4$).

Because the interactions in the band basis are linear combinations of Hubbard and Hund interactions in the orbital basis, weighted with
 orbital coherence factors, the sign of $u_1$ and $u_3$ depends on the interplay between intra-orbital and inter-orbital interactions~\cite{kontani,raghu}.
  The interaction $u_3$ contributes to the superconducting channel, and for an $s^{+-}$ gap structure must be repulsive. The sign of $u_1$ is a priori unknown. In most microscopic studies it comes out positive (repulsive), but in principle it can also be negative (attractive).  We do not assume a particular sign
      of $u_1$ and consider first a case where $u_1$ is positive and then when it is negative. For the first case we show that a resonance can only originate from the particle-hole channel. For negative $u_1$ the $\pi$ channel can produce collective modes as well, and we show that in general the resonant mode is a mix between spin exciton and a $\pi$-resonance.

\section{Susceptibilities and RPA}
\label{chi_rpa}

We focus on susceptibilities at antiferromagnetic momentum $%
\mathbf{Q}$ which separates the centers of hole and electron pockets. Following similar work done on the cuprates\cite{norman}, we
define spin and $\pi$  operators as
\begin{align}
S^{z}(\mathbf{Q})& =\frac{1}{\sqrt{N}}\sum_{\mathbf{k}}\left[ c_{\mathbf{k}%
\alpha }^{\dagger }\sigma _{\alpha \beta }^{z}f_{\mathbf{k}+\mathbf{Q}\beta
}+f_{\mathbf{k}+\mathbf{Q}\alpha }^{\dagger }\sigma _{\alpha \beta }^{z}c_{%
\mathbf{k}\beta }\right] , \\
\pi (\mathbf{Q})& =\frac{1}{\sqrt{N}}\sum_{\mathbf{k}}\left[ c_{\mathbf{k}%
\alpha }\sigma _{\alpha \beta }^{x}f_{\mathbf{Q}-\mathbf{k}\beta }\right] .
\end{align}
 To make a closer connection to\cite{norman}, the operator $\pi$ can
be equivalently defined as $\pi= \frac{1}{\sqrt{N}} \sum_{\mathbf{k}}\left[
g_\mathbf{k} a_{\mathbf{k} \alpha} \sigma^x_{\alpha \beta} a_{\mathbf{Q}-%
\mathbf{k} \beta} \right]$, with $|g_\mathbf{k}|=1/2$ and the sign of $g_%
\mathbf{k}$ is chosen so that it is positive near the hole FS and negative
near the electron FS ($g_\mathbf{k}=-g_{\mathbf{Q}-\mathbf{k}}$).

For notational convenience we split $S^z$ into two operators
such that $S^z=S_c+S_f$, where
\begin{align}
S_c(\mathbf{Q}) &= \frac{1}{\sqrt{N}} \sum_{\mathbf{k}} c^\dagger_{\mathbf{k}
\alpha} \sigma^z_{\alpha \beta} f_{\mathbf{k}+\mathbf{Q} \beta}, \\
S_f(\mathbf{Q}) &= \frac{1}{\sqrt{N}} \sum_{\mathbf{k}} f^\dagger_{\mathbf{k}%
+\mathbf{Q} \alpha} \sigma^z_{\alpha \beta} c_{\mathbf{k} \beta}.
\end{align}

We now define the susceptibilities $\chi_{a b}(\Omega_m)$ in terms of Matsubara frequencies as
\begin{equation}
	\chi_{a b}(\Omega_m)= \int_0^{1/T}\mathrm{d}\tau' e^{i\Omega_m \tau'}\left\langle T_\tau A_a(\tau') A^\dagger_b(0)\right\rangle,
\end{equation}
where $A_a=(S_c,S_f,\pi,\pi^{\dag})_a$. The actual spin
susceptibility is given by $\chi_S=\chi_{11}+\chi_{12}+\chi_{21}+\chi_{22}$.

The bare susceptibilities $\chi _{ab}^{0}$ can be calculated in the usual way in terms of Green's functions and are given by bubbles made out of $c$ and $f$-fermions, with different Pauli matrices in the vertices.
At $T=0$ and after performing analytic continuation to real frequency space,
the (retarded) susceptibilities have the following form:
\begin{align}
\chi^0_{a b}(\omega) =\frac{2}{N}\sum_\mathbf{k} \Bigg[&-\frac{A_{ab}(%
\mathbf{k})}{\omega -E^c_\mathbf{k} -E^f_{\mathbf{k}+\mathbf{Q}}+i\gamma}
\notag \\
&+\frac{B_{ab}(\mathbf{k})}{\omega +E^c_\mathbf{k} +E^f_{\mathbf{k}+\mathbf{Q%
}}+i\gamma}\Bigg],
\end{align}
where $\chi^0_{a b}$, $A_{ab}$, and $B_{ab}$ are symmetric matrices. The expressions for $%
A_{ab}$ and $B_{ab}$ are presented in Table \ref{tab:Coefficients} in terms of
coherence factors which are given by:
\begin{align}
u_\mathbf{k}^c &=\sqrt{\frac{1}{2}\left(1+\frac{\xi_{\mathbf{k}}^c}{E^c_%
\mathbf{k}} \right)}, \\
v_\mathbf{k}^c &=\sqrt{\frac{1}{2}\left(1-\frac{\xi_{\mathbf{k}}^c}{E^c_%
\mathbf{k}} \right)} \sgn \Delta^c,
\end{align}
and similar expressions for $u^f_{\mathbf{k}^\prime}$ and $v^f_{\mathbf{k}%
^\prime}$.

To obtain the full susceptibilities $\chi _{ab}$ we used the generalized RPA approach. Within this approach
\begin{equation}
\chi _{ab}=(1-\chi ^{0}V)_{ac}^{-1}\chi _{cb}^{0},
\end{equation}%
where the sum over repeated indices is implied and $V$ is given by
\begin{equation}
V=\frac{1}{2}%
\begin{pmatrix}
u_{1} & u_{3} & 0 & 0 \\
u_{3} & u_{1} & 0 & 0 \\
0 & 0 & -u_{1} & 0 \\
0 & 0 & 0 & -u_{1}%
\end{pmatrix}%
.
\end{equation}

\begin{table}[tbp]
\centering
\begin{tabular}{cccc}
\hline
$a$ & $b$ & $A_{ab}(\mathbf{k})$ & $B_{ab}(\mathbf{k})$ \\ \hline
1 & 1 & $(v^c_\mathbf{k} u^f_{\mathbf{k}^{\prime }})^2$ & $(u^c_\mathbf{k}
v^f_{\mathbf{k}^{\prime }})^2$ \\
1 & 2 & $-u^c_\mathbf{k} v^c_\mathbf{k} u^f_{\mathbf{k}^{\prime }} v^f_{%
\mathbf{k}^{\prime }}$ & $-u^c_\mathbf{k} v^c_\mathbf{k} u^f_{\mathbf{k}%
^{\prime }} v^f_{\mathbf{k}^{\prime }}$ \\
1 & 3 & $u^c_\mathbf{k} v^c_\mathbf{k} (u^f_{\mathbf{k}^{\prime }})^2$ & $%
-u^c_\mathbf{k} v^c_\mathbf{k} (v^f_{\mathbf{k}^{\prime }})^2$ \\
1 & 4 & $(v^c_{\mathbf{k}})^2 u^f_{\mathbf{k}^{\prime }} v^f_{\mathbf{k}%
^{\prime }}$ & $-(u^c_{\mathbf{k}})^2 u^f_{\mathbf{k}^{\prime }} v^f_{%
\mathbf{k}^{\prime }}$ \\
2 & 2 & $(u^c_\mathbf{k} v^f_{\mathbf{k}^{\prime }})^2$ & $(v^c_\mathbf{k}
u^f_{\mathbf{k}^{\prime }})^2$ \\
2 & 3 & $-(u^c_{\mathbf{k}})^2 u^f_{\mathbf{k}^{\prime }} v^f_{\mathbf{k}%
^{\prime }}$ & $(v^c_{\mathbf{k}})^2 u^f_{\mathbf{k}^{\prime }} v^f_{\mathbf{%
k}^{\prime }}$ \\
2 & 4 & $-u^c_\mathbf{k} v^c_\mathbf{k} (v^f_{\mathbf{k}^{\prime }})^2$ & $%
u^c_\mathbf{k} v^c_\mathbf{k} (u^f_{\mathbf{k}^{\prime }})^2$ \\
3 & 3 & $(u^c_\mathbf{k} u^f_{\mathbf{k}^{\prime }})^2$ & $(v^c_\mathbf{k}
v^f_{\mathbf{k}^{\prime }})^2$ \\
3 & 4 & $u^c_\mathbf{k} v^c_\mathbf{k} u^f_{\mathbf{k}^{\prime }} v^f_{%
\mathbf{k}^{\prime }}$ & $u^c_\mathbf{k} v^c_\mathbf{k} u^f_{\mathbf{k}%
^{\prime }} v^f_{\mathbf{k}^{\prime }}$ \\
4 & 4 & $(v^c_\mathbf{k} v^f_{\mathbf{k}^{\prime }})^2$ & $(u^c_\mathbf{k}
u^f_{\mathbf{k}^{\prime }})^2$ \\
&  &  &
\end{tabular}
\caption{Coefficients of bare susceptibilities. Note: $\mathbf{k}^{\prime }=%
\mathbf{k}+\mathbf{Q}$.}
\label{tab:Coefficients}
\end{table}

 The solution for the full spin susceptibility can be written in a simpler
form once we note that the matrix $\chi _{ab}^{0}$ has additional symmetry.
Indeed, the functions $A_{ab}(\mathbf{k})$ and $B_{ab}(\mathbf{k})$ can be separated into parts
that are even or odd with respect to $\xi _{\mathbf{k}}^{c}$
 and $\xi _{\mathbf{k}}^{f}$. If the momentum sums are evaluated only near the FSs,
where the integration region can be chosen to be symmetric with respect to
positive and negative values of $\xi _{\mathbf{k}}^{c}$ and $\xi _{\mathbf{k}%
}^{f}$,
 then the odd parts cancel out and $\chi _{ab}^{0}$ acquires the following
form:
\begin{equation}
\chi ^{0}=%
\begin{pmatrix}
a & b & c & -d \\
b & a & d & -c \\
c & d & e & f \\
-d & -c & f & e%
\end{pmatrix}%
\end{equation}

\begin{figure}[htb]
\centering
\includegraphics[width=.45\textwidth]{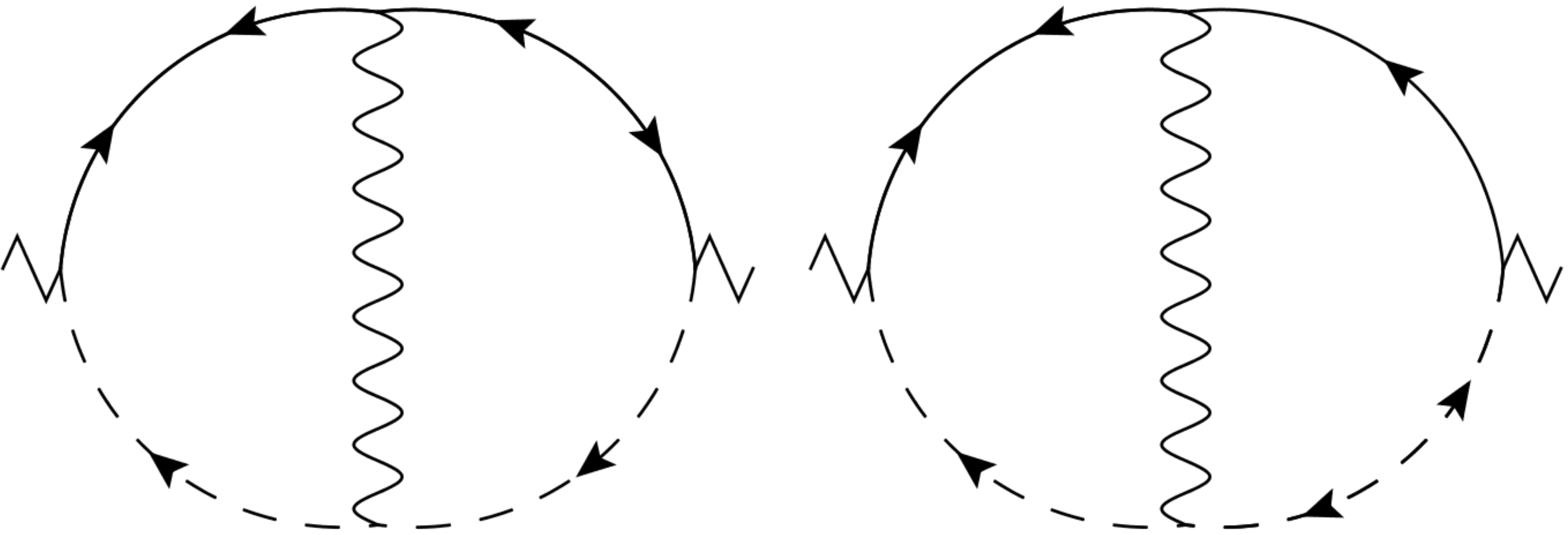}
\caption{Some mixed-channel contributions to the spin susceptibility and Raman scattering. The solid lines represent $f$ fermions and the dashed lines $c$ fermions, corresponding to quasiparticles from the electron and hole pockets, respectively.}
\label{fig:diagrams}
\end{figure}

If we rotate the basis of operators from $(S_c,S_f,\pi,\pi^{\dagger})$ to $%
(S_c+S_f,\pi-\pi^{\dagger},S_c-S_f,\pi+\pi^{\dagger})$ we find that both $V
$ and $\chi^0$ become block diagonal, so the first $2 \times 2$ system
decouples from the second one. In this case, the solution for the subset $%
(S_c+S_f,\pi-\pi^{\dagger})$ takes the simple form

\begin{align}
\chi_S= & \frac{\chi^0_S+\delta \chi^0_S}{1-u_S (\chi^0_S+\delta \chi^0_S)},
\label{eq:RPAchiS} \\
\chi_\pi= & \frac{\chi^0_\pi+\delta \chi^0_\pi}{1-u_\pi (\chi^0_\pi+\delta
\chi^0_\pi)},
\end{align}
where
\begin{align}
\delta \chi^0_S= & \frac{u_\pi}{1-u_\pi \chi^0_\pi}\left(\chi^0_{S\pi}%
\right)^2,  \label{eq:delta_chi0S} \\
\delta \chi^0_\pi= & \frac{u_S}{1-u_S \chi^0_S}\left(\chi^0_{S\pi}\right)^2.
\end{align}
and $u_S=(u_1+u_3)/4$ and $u_\pi=-u_1/4$. The bare susceptibilities in this
basis are given by
\begin{align}
\chi^0_S &=
\chi^0_{11}+\chi^0_{12}+\chi^0_{21}+\chi^0_{22}=2(\chi^0_{11}+\chi^0_{12}),
\\
\chi^0_\pi &=
\chi^0_{33}-\chi^0_{34}-\chi^0_{43}+\chi^0_{44}=2(\chi^0_{33}-\chi^0_{34}),
\\
\chi^0_{S\pi} &=
\chi^0_{13}-\chi^0_{14}+\chi^0_{23}-\chi^0_{24}=2(\chi^0_{13}-\chi^0_{14}).
\end{align}
The first two bare susceptibilities contain contributions of products of two normal Green's functions and of two anomalous Green's functions, while
 $\chi^0_{S\pi}$ is composed of one normal and one anomalous Green's function (see  Fig. \ref{fig:diagrams} for some contributions of $\chi^0_{S\pi}$ to the spin response function).

We obtained expressions for the real parts of the bare susceptibilities by replacing the momentum sums by integrals and evaluating them within an energy range from $-\Lambda$ to $\Lambda$ about the FSs.  They are valid in the range $0<\omega<2\Delta$ in the limit when the broadening $\gamma\rightarrow 0$.
\begin{align}
\mathrm{Re} \chi^0_S (\omega) =& L+\frac{\omega^2}{4\Delta^2} \mathrm{Re} \chi^0_\pi (\omega), \label{eq:Rechis}\\
\mathrm{Re} \chi^0_\pi (\omega) =& \frac{m}{\pi} \frac{4\Delta^2}{\omega\sqrt{4\Delta^2-\omega^2}}\arctan \left( \frac{\omega}{\sqrt{4\Delta^2-\omega^2}}\right),\\
\mathrm{Re} \chi^0_{S\pi} (\omega) =& \frac{\omega}{2\Delta} \mathrm{Re} \chi^0_\pi (\omega),\label{eq:Rechisp}
\end{align}
where $L=\frac{m}{\pi} \log (2\Lambda/\Delta)$ and we have neglected terms of order $(\Delta/\Lambda)^2$.

\begin{figure}[tb]
\centering
\includegraphics[width=0.5\textwidth]{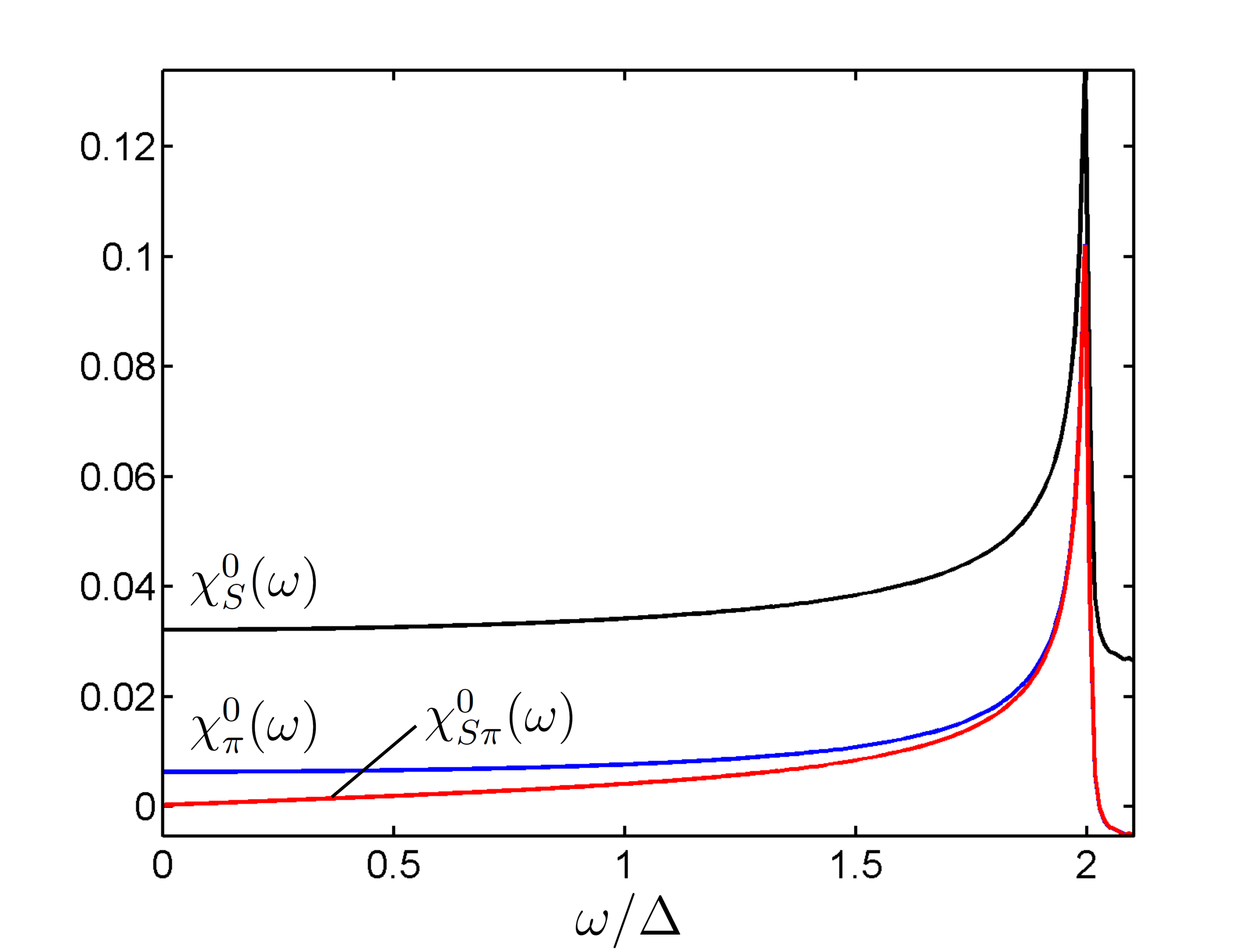}
\caption{Real part of the bare spin susceptibilities ($T=0$, units of $\Delta^{-1}$) in the case of
perfectly nested circular pockets ($m\equiv m_c=m_f=\frac{1}{100\Delta}$).
In all numerical calculations $\protect\mu=10 \Delta$
and we include a finite broadening $\protect\gamma=\Delta/200$ when
evaluating momentum integrals.}
\label{fig:barechi}
\end{figure}

 In Fig. \ref{fig:barechi} we present the results of numerical calculations of the bare
susceptibilities in the case of perfectly nested FSs ($\mu_c = \mu_f \equiv \mu$). The susceptibility $\chi^0_S$ would diverge at $\omega=0$ in the absence of superconductivity, but becomes finite at a finite $\Delta$. Conversely, $\chi^0_\pi$ at $\omega=0$ would be zero in the absence of superconductivity, but becomes non-zero because of $\Delta$.  Note that all three susceptibilities  monotonically increase with frequency in the domain $0\leq \omega\leq 2\Delta$ and diverge at $\omega=2\Delta$. The imaginary parts of the three bare susceptibilities (not shown in the plot) are infinitesimally small and undergo a discontinuous jump at $\omega=2\Delta$.
If we make the electron pocket elliptical, the divergence in the real part is replaced by a local maximum.

We see that $\chi^0_S (\omega)$ and $\chi^0_\pi(\omega)$ have finite value at $\omega =0$. We recall that, in the absence of superconductivity, $\chi_S (0)$ would diverge and $\chi_\pi (0)$ would vanish at perfect nesting.  At $\omega =2\Delta$ both bare susceptibilities diverge.  The cross-susceptibility $\chi^0_{S\pi} (\omega)$ vanishes at $\omega =0$ simply because it is composed from one normal and one anomalous Green's function but rapidly increases with $\omega$  and becomes comparable to $\chi_S (\omega)$ and $\chi_\pi (\omega)$ at $\omega \leq 2 \Delta$.

The cross-susceptibility between particle-hole and particle-particle channels has been recently analyzed for an $s^{+-}$ superconductor in the context of Raman scattering~\cite{raman}. There, it was computed in the charge channel and was found to be very small due to near-cancellation between contributions from Fermi surfaces with
 plus and minus values of the superconducting gap.  In our case we found that the contributions from hole and electron FSs add up rather than cancel.
 The difference is that in Raman scattering the side vertices in the susceptibility bubble have the spin structure given by $\delta_{\alpha, \beta}$, while in our case the spin structure is, say,  $\sigma^z_{\alpha, \beta}$. For an $s^{+-}$ gap,  Raman bubbles from hole and electron pocket have the same vertex structure but differ in the sign of the anomalous Green's function, hence the two contributions to cross-susceptibility
  have opposite signs, resulting in a cancellation that is complete in the case of perfect nesting and near-complete in the case of one circular and one elliptical pocket. This cancellation does not occur in our case because the $\sigma^z$ structure of the side vertices additionally flips the sign of one of the two diagrams, and the contributions to $\chi^0_{S\pi} (\omega)$ from hole and electron FSs add constructively.

We next note that the terms $\delta \chi _{S}^{0}$ and $\delta \chi _{\pi}^{0}$ are precisely what is neglected
when the particle-particle channel is not included in the calculation of the
spin susceptibility.  Setting these terms to  zero reduces the expressions for the full $\chi_S$ and $\chi_\pi$ to the usual RPA
results
\begin{equation}
\chi _{S}=\frac{\chi _{S}^{0}}{1-u_{S}\chi _{S}^{0}}, \quad\quad
\chi _{\pi}=\frac{\chi _{\pi}^{0}}{1-u_{\pi}\chi _{\pi}^{0}}.\label{eq:chi_fullform}
\end{equation}%
The effect of coupling the two channels can be seen more clearly by
substituting (\ref{eq:delta_chi0S}) into (\ref{eq:RPAchiS}), which yields
\begin{align}
\chi _{S}=&\frac{\chi _{S}^{0}(1-u_{\pi }\chi _{\pi }^{0})+u_{\pi }(\chi
_{S\pi }^{0})^{2}}{(1-u_{S}\chi _{S}^{0})(1-u_{\pi }\chi _{\pi
}^{0})-u_{S}u_{\pi }(\chi _{S\pi }^{0})^{2}} \label{eq:fullchi} \\
 =& \frac{1}{u_S} \left(-1 + \frac{1-u_{\pi }\chi _{\pi }^{0}}{(1-u_{S}\chi _{S}^{0})(1-u_{\pi }\chi _{\pi
}^{0})-u_{S}u_{\pi }(\chi _{S\pi }^{0})^{2}}\right) \nonumber
\end{align}%
The positions of resonance peaks are given by the zeroes of the denominator in this equation and we can see that the particle-hole and particle-particle channels are coupled through the mixed-channel susceptibility $\chi^0_{S\pi}(\omega)$.

\section{The results}
\label{results}

\subsection{Purely repulsive interaction, $u_1 >0$, $u_3 >0$.}
\begin{figure}[htb]
\centering
\includegraphics[width=.45\textwidth]{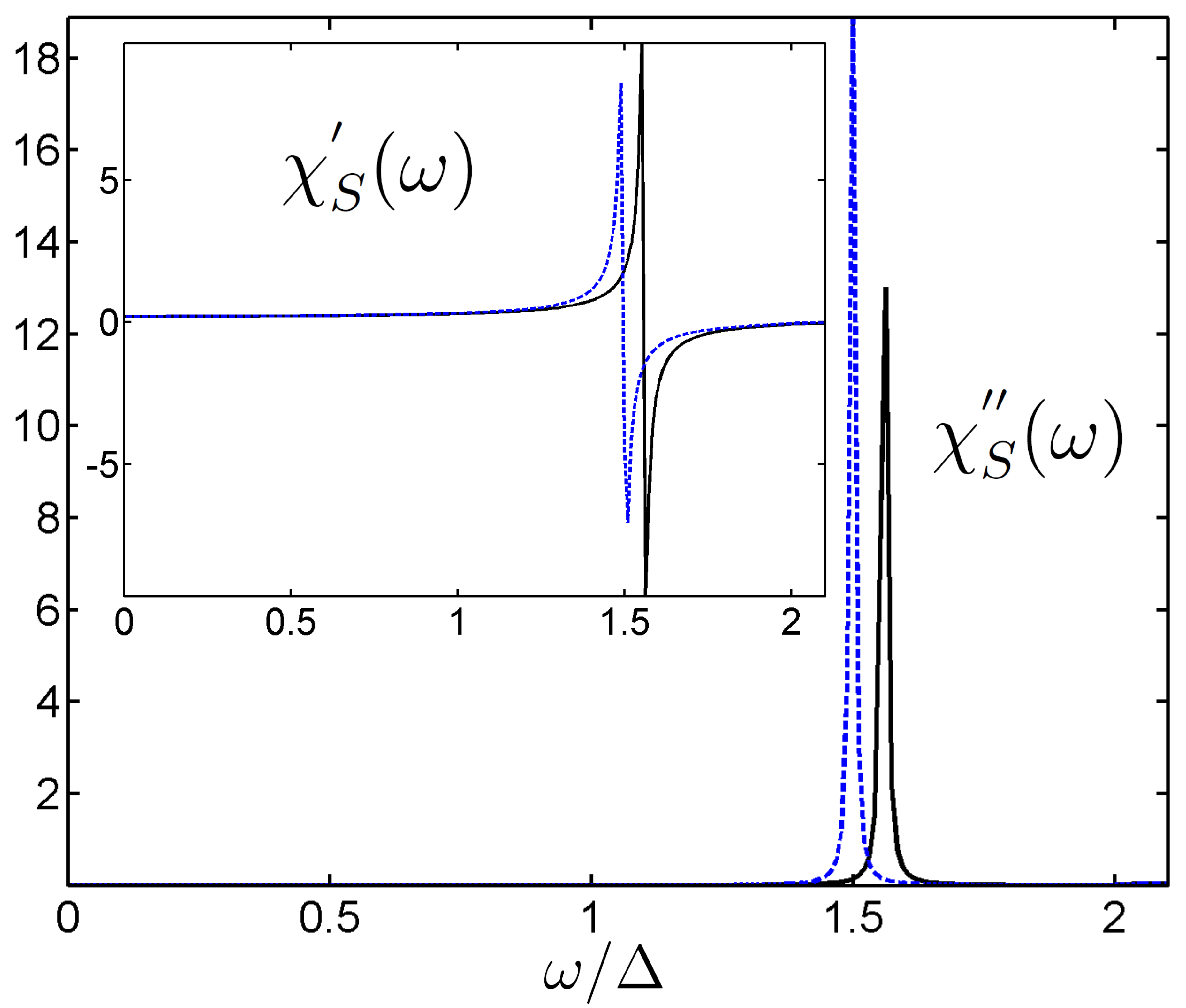}
\caption{Imaginary part (main plot) and real part (insert) of the full spin susceptibility ($T=0$, units of $\Delta^{-1}$) in the case of repulsive interactions. In this plot $u_S=26\Delta$ and $u_\protect\pi=-13\Delta$.
The solid, black line indicates the full calculation. For comparison, the dashed, blue line indicates a calculation that neglects particle-particle contributions.}
\label{fig:fullchi_cc}
\end{figure}

For repulsive interactions $u_{S}>|u_\pi|>0$ and
$u_{\pi }<0$. In the absence of $\chi^0_{S\pi}$  the resonance in $\chi_S$ is present for any $u_S$  because the bare susceptibility $\chi _{S}^{0}$ is positive and  diverges at $\omega =2\Delta $,  hence the
equation $1-u_{S}\mathrm{Re}\chi _{S}^{0}(\omega )=0$ has a solution for $0<u_{S}<(\mathrm{Re} \chi _{S}^{0}(0))^{-1}$. In contrast, the fact that
$\mathrm{Re} \chi _{\pi }^{0}>0$ means that no resonance originates from this
channel.
When $\chi^0_{S\pi}$ is included, we found in our numerical calculations  that  the effect of the
particle-particle channel is that  the peak in the imaginary part of the full
susceptibility is shifted to a higher frequency (since $\mathrm{Re} \{u_{S}u_{\pi
}(\chi _{S\pi }^{0})^{2}\}<0$). We show representative behavior of real and imaginary parts of the full spin susceptibility in  Fig. \ref{fig:fullchi_cc}

This result is also obtained when we consider an elliptical electron pocket, except that there is a minimum value for $u_S$ below which no resonance is observed. This is due to the fact that the bare susceptibilities have a local maximum instead of a divergence at $\omega=2\Delta$.

\subsection{Partially attractive interaction, $u_1 <0, u_3 >0$.}

We now consider an alternative case where the density-density interaction $u_1 <0$, hence $u_\pi = - u_1/4 >0$. We still assume $u_3 > |u_1|$ such that $u_S >0$.
 For positive $u_\pi$, the $\pi$ channel can acquire a resonance on its own since the equation $1-u_{\pi}\mathrm{Re}\chi _{\pi}^{0}(\omega)=0$ necessarily
 has a solution at a frequency between $0$ and $2\Delta$ if $u_{\pi}<(\mathrm{Re} \chi _{\pi}^{0}(0))^{-1}$. We assume that this inequality holds together with
  $u_{S}<(\mathrm{Re} \chi _{S}^{0}(0))^{-1}$. If any of these two conditions are not satisfied, the system becomes unstable either towards $\pi$ superconductivity
	with  total momentum of a Cooper pair ${\bf Q}$ or towards magnetic order. In both cases, the analysis of the spin susceptibility has to be modified to include
the new condensates.
 If the spin-exciton and $\pi$  channels were not coupled (i.e., if $\chi^0_{S\pi}(\omega)$ was absent), we would find resonances in the spin and $\pi$ channels at frequencies
$\omega_{S}$ and $\omega _{\pi }$, respectively, set by $1-u_{S}\mathrm{Re}\chi_{S}^{0}(\omega _{S})=0$ and $1-u_{\pi }\mathrm{Re}\chi _{\pi }^{0}(\omega _{\pi })=0$. This suggests the possibility that there may be two resonance peaks in the full spin susceptibility $\chi_S (Q, \omega)$ once we restore the coupling $\chi^0_{S\pi} (\omega)$.

\begin{figure}[htb]
\centering
\includegraphics[width=.45\textwidth]{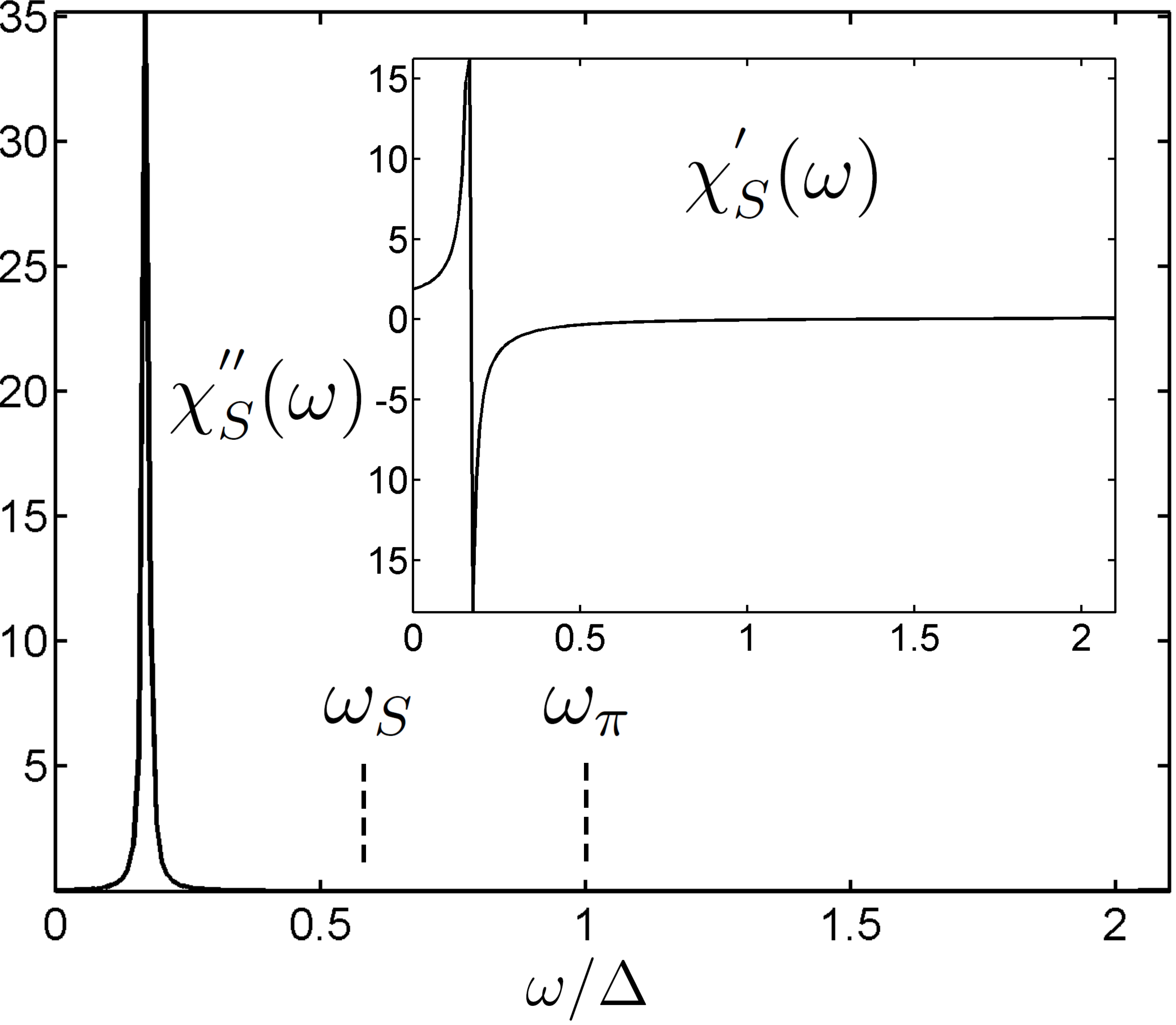}
\caption{Imaginary part (main plot) and real part (insert) of the full spin susceptibility ($T=0$, units of $\Delta^{-1}$) in the case of $u_S=30.5\Delta$ and $u_\protect\pi=130\Delta$. In the absence of coupling we would observe resonance peaks in the spin and $\pi$ channels at frequencies $\omega_S\approx 0.57\Delta$ and $\omega_\pi\approx 1.0 \Delta$, respectively, indicated on the plot.}
\label{fig:chi_attractive}
\end{figure}

However, we found that for all values of $u_S$ and $u_\pi$
 for which the pure $s^{+-}$ state is stable, there is only a single peak  in
  the spin susceptibility at a frequency lower than both $\omega_S$ and $\omega_\pi$. We show representative behavior in Fig. \ref{fig:chi_attractive}.
   The existence of a single peak is  due to the fact that $\chi^0_{S\pi}$ is small at small frequencies, hence it does not prevent the increase of the real part of the spin susceptibility with increasing $\omega$ (see insert in Fig. \ref{fig:chi_attractive}) and only shifts the position of the lower pole ($\omega_S$ or $\omega_\pi$) to a smaller value $\omega_{res}$.  At the same time, at higher frequencies,  $\chi^0_{S\pi}$  is no longer small relative to the other bare susceptibilities
     $\chi^0_S$ and $\chi^0_\pi$.
       As a result, the denominator in $\chi_S(\bf Q, \omega)$ in (\ref{eq:fullchi}) passes through zero at $\omega =\omega_{res}$ and then remains negative all the way up to $\omega = 2\Delta$ and does not cross zero for the second time.

To better understand this, we artificially add a factor $\epsilon$ to $\chi^0_{S\pi}$ and consider how the solutions evolve as we progressively increase $\epsilon$ between $0$ and $1$. At small $\epsilon$, the two solutions obviously survive and just further split from each other --  the peak that was at a higher frequency  shifts to a higher frequency and the other peak shifts to to a lower frequency.  As $\epsilon$ increases, the peak at a higher frequency rapidly moves
 towards $2\Delta$. If we keep $\mathrm{Im} \chi^0_{ab}$ strictly zero,  this peak survives up to $\epsilon =1$ with exponentially vanishing amplitude. If, however, we
  keep a small but finite fermionic damping in the computations of $\chi^0_{ab}$, we find that the functions $\chi^0_{ab} (\omega)$ increase but do not diverge at $2\Delta$. In this situation, the higher frequency peak in $\chi_S (\bf Q, \omega)$ vanishes already at some $\epsilon <1$.

  We also considered the evolution of the two-peak solution with $\epsilon$ in a different way: we postulated that the two peaks should be at
   $\omega_{res,1}$ and $\omega_{res,2}$, both below $2\Delta$ and solved the set of equations for $u_S$ and $u_\pi$ which would correspond to such a solution.   At small $\epsilon$ we indeed found  some real $u_S$ and $u_\pi$ which satisfy ``boundary conditions'' $u_S \chi^0_S <1$ and $u_\pi \chi^0_\pi <1$.
    However this holds only up to some $\epsilon_{cr}$. At higher $\epsilon$  the solutions for $u_s$ and $u_\pi$ become complex, which implies that the two-peak solution is no longer possible.  At even higher $\epsilon$ real solutions for $u_s$ and $u_\pi$ reappear, but they do not satisfy the boundary conditions.
     We searched for a range of $\omega_{res,1}$ and $\omega_{res,2}$ and for all values that we tested we found $\epsilon_{cr} <1$, i.e., again there is only a single peak for the actual case of $\epsilon =1$.

 Another way to see that there is only one peak in the full $\chi _S$ is to substitute the expressions for the real parts of the susceptibilities, Eqs. (\ref{eq:Rechis})-(\ref{eq:Rechisp}), into the denominator of Eq. (\ref{eq:fullchi}) and express the real part of the term $D= (1-u_{S}\chi _{S}^{0})(1-u_{\pi }\chi _{\pi}^{0})-u_{S}u_{\pi }(\chi _{S\pi }^{0})^{2}$ via $\chi^0_\pi = \chi^0_\pi (\omega)$. We obtain

\begin{equation}
	\mathrm{Re} D=(1-u_S L) -\left[u_S\frac{\omega^2}{4\Delta^2}+u_\pi(1-u_S L) \right] \mathrm{Re} [\chi^0_\pi (\omega)].
\label{nn_1}
\end{equation}
 Because $(1-u_S L)$ and $(1 - u_\pi \chi^0_\pi (0))$ are required to be positive for the stability of the paramagnetic state,
  at zero frequency, $D$ is surely positive.  At finite $\omega$,
 the first term in (\ref{nn_1})
  is positive, while the second one is negative and its magnitude monotonically increases with increasing $\omega$.  As a result,
   the denominator crosses zero only once, at some $\omega < 2\Delta$.

  The single resonance peak is a mixture of a spin-exciton and $\pi$-resonance and for the representative case shown in Fig. \ref{fig:chi_attractive}
   its energy is smaller that that of spin-exciton and a $\pi$-resonance. This implies that, when both channels are attractive, the coupling between the two plays substantial role in determining the position of the true resonance. From this perspective, the resonance at $u_1 <0$ can,  at least partly, be viewed as a  plasmon. A somewhat similar result has been earlier obtained in the analysis of the resonance in the cuprates in the parameter range where $\pi-$resonance is allowed~\cite{norman,hao,macdonald}.

\section{Conclusions}
\label{concl}
We have studied the spin resonance at antiferromagnetic momentum $(\pi,\pi)$ in an $s^{+-}$ superconducting state of
 FeSCs by including contribution from the particle-particle channel, which in the superconducting state gets mixed with the particle-hole channel.
   We have shown that for purely repulsive interactions the inclusion of this channel does not qualitatively change the
spin resonance, which remains predominantly spin-exciton and only slightly shifts to higher frequencies.  For attractive density-density interaction, when both spin-exciton resonance in the particle-hole channel and $\pi$-resonance in the particle-particle channel are allowed, we found that strong coupling
 between the two channels destroys the two-peak structure and only one peak survives, whose frequency is smaller than would be that of a spin-exciton and $\pi-$resonance in the absence of the coupling.  We argued that strong coupling between the particle-hole and particle-particle channels is peculiar to the spin
 susceptibility, while for the charge susceptibility, which, e.g., is relevant for Raman scattering,  the coupling is much smaller.

We acknowledge useful discussions with R. Fernandes, S. Maiti,  and Y. Wang.
The work by A. H. and A.V.C.
is supported by the DOE grant DE-FG02-ER46900.
P. W. thanks the Department of Physics at the University of Wisconsin-Madison for hospitality during a stay as a visiting professor. P. W. also acknowledges partial support through the DFG research unit ``Quantum phase transitions''.


\begin{thebibliography}{10}

\bibitem{rossat} J. Rossat-Mignod, L.P. Regnault, C. Vettier, P. Bourges, P.
Burlet, J. Bossy, J.Y. Henry, and G. Lapertot, Physica C \textbf{185-189},
86 (1991);  H. A. Mook, M. Yethitaj, G. Aeppli, T. E. Mason, and T.
Armstrong, Phys. Rev. Lett. \textbf{70}, 3490 (1993); H. F. Fong, B. Keimer, P. W. Anderson, D. Reznik, F. Dogan,
and I. A. Aksay, Phys. Rev. Lett. \textbf{75}, 316 (1995); H.F. Fong et al, Nature {\bf 398}, 588 (1999); Ph. Bourges et al, Science 288 1234 (2000);
 H. He et al, Science {\bf 295}, 1045 (2002); G. Yu, Y. Li, E. M. Motoyama, and M. Greven, Nat. Phys.
\textbf{5}, 873 (2009).
\bibitem{heavy_fermion}
 N.K. Sato, N. Aso, K. Miyake, R. Shiina, P. Thalmeier, G. Varelogiannis, C. Geibel, F. Steglich, P. Fulde, and T. Komatsubara,
    Nature \textbf{410}, 340 (2001);
 C. Stock, C. Broholm, J. Hudis, H.J. Kang, and C. Petrovic,
    Phys. Rev. Lett. \textbf{100}, 087001 (2008).
\bibitem{claus} A. D. Christianson, E. A. Goremychkin, R. Osborn, S.
Rosenkranz, M. D. Lumsden, C. D. Malliakas, I. S. Todorov, H. Claus, D. Y.
Chung, M. G. Kanatzidis, R. I. Bewley, and T. Guidi, Nature (London) \textbf{%
456}, 930 (2008); M. D. Lumsden, A. D. Christianson, D. Parshall, M. B. Stone,
S. E. Nagler, G. J. MacDougall, H. A. Mook, K. Lokshin, T. McGuire, A. S.
Sefat, R. Jin, B. C. Sales, and D. Mandrus, Phys. Rev. Lett. \textbf{102},
107005 (2009); S. Chi, A. Schneidewind, J. Zhao, L. W. Harriger, L. Li, Y.
Luo, G. Cao, Z. Xu, M. Loewenhaupt, J. Hu, and P. Dai, Phys. Rev. Lett.
\textbf{102}, 107006 (2009); S. Li, Y. Chen, S. Chang, J.W. Lynn, L. Li,
Y. Luo, G. Cao, Z. Xu, and P. Dai, Phys. Rev. B {\bf 79}, 174527
(2009); D.S. Inosov, J.T. Park, P. Bourges, D.L. Sun, Y. Sidis, A. Schneidewind, K. Hradil, D. Haug, C. T. Lin, B. Keimer, V. Hinkov, Nature Phys. {\bf 6}, 178 (2010);
J.-P. Castellan, S. Rosenkranz, E. A. Goremychkin, D.
Y. Chung, I. S. Todorov, M. G. Kanatzidis, I. Eremin,
J. Knolle, A. V. Chubukov, S. Maiti, M. R. Norman, F.
Weber, H. Claus, T. Guidi, R. I. Bewley, and R. Osborn,
arXiv:1106.0771;
Y. Qiu, W. Bao, Y. Zhao, C. Broholm, V. Stanev, Z.
Tesanovic, Y. C. Gasparovic, S. Chang, J. Hu, B. Qian, M.
Fang, and Z. Mao, Phys. Rev. Lett. 103, 067008 (2009);
D.N. Argyriou, A. Hiess, A. Akbari, I. Eremin, M.M. Korshunov,
J. Hu, B. Qian, Z. Mao, Y. Qiu, C. Broholm, and
W. Bao, Phys. Rev. B 81, 220503(R) (2010).
\bibitem{scalapino} N. Bulut, and D. J. Scalapino, Phys. Rev. B \textbf{47}%
, 3419 (1993); I. Mazin and V. Yakovenko, Phys. Rev. Lett. 75 4134 (1995); D. Z. Liu, Y. Zha, and K. Levin, Phys. Rev. Lett. 75 4130 (1995);
 N. Bulut and D. Scalapino, Phys. Rev. B {\bf 53}, 5149 (1996); A. J. Millis and H. Monien, Phys. Rev. B, {\bf 54}, 16172 (1996); D. Manske, I. Eremin, and K.~H. Bennemann, Phys. Rev. B {\bf 63}, 054517 (2001);
M.R. Norman, Phys. Rev. B {\bf 61}, 14751
(2000); {\it ibid} {\bf 63}, 092509 (2001);  A. Chubukov,  B.
Janko and O. Tchernyshov, Phys. Rev. B {\bf 63}, 180507(R) (2001);  A. Abanov, A. Chubukov, and J. Schmalian,
Journal of Electron Spectroscopy and Related Phenomena 117, 129-151 (2001);
F. Onufrieva and P. Pfeuty, Phys. Rev. B 65, 054515 (2002);
 I. Eremin, D.K. Morr, A.V. Chubukov, K.H. Bennemann,
and M.R. Norman, Phys. Rev. Lett. 94, 147001 (2005); Ar Abanov, A. Chubukov, and M. Norman, Phys. Rev. B 78, 220507 (2008).
\bibitem{norman} O. Tchernyshyov, M. R. Norman, and A. V. Chubukov, Phys.
Rev. B \textbf{63}, 144507 (2001).
\bibitem{res_fe}  M.M. Korshunov, and I. Eremin, Phys. Rev. B {\bf 78}, 140509(R) (2008); T.A. Maier, and D.J. Scalapino, Phys. Rev. B {\bf 78}, 020514(R)
(2008); T. A. Maier, S. Graser, D. J. Scalapino, and P. Hirschfeld, Phys. Rev. B {\bf 79}, 134520 (2009);  J. Zhang, R. Sknepnek, and J. Schmalian, Phys. Rev. B
\textbf{82}, 134527 (2010);
S. Maiti, J. Knolle, I. Eremin, and A.V. Chubukov,  Phys.
Rev. B \textbf{84}, 144524 (2011).
\bibitem{scalapino12} D. J. Scalapino, Rev. Mod. Phys. \textbf{84}, 1383
(2012).
\bibitem{brinck} J. Brinckmann, and P. A. Lee, Phys. Rev. Lett. \textbf{82},
2915 (1999); K. Seo, C. Fang, B.A. Bernevig, and J. Hu, Phys. Rev. B {\bf 79}, 235207 (2009).
\bibitem{eschrig} for a review see  M. Eschrig, Adv. Phys. {\bf 55}, 47 (2006).
\bibitem{morr} D. Morr and D. Pines, Phys. Rev. Lett. \textit{81}, 1086 (1998).
\bibitem{gorkov} A.V. Chubukov and L.P. Gorkov,  Phys. Rev. Lett. 101,
147004 (2008).
\bibitem{zhang} E. Demler, and S.-C. Zhang, Phys. Rev. Lett. \textbf{75, }%
4126 (1995); E. Demler, H. Kohno, and S.-C. Zhang, Phys. Rev. B \textbf{%
58}, 5719 (1998).
\bibitem{kiv_fra} E. Fradkin and S. A. Kivelson,  Nature Physics {\bf 8}, 865-866 (2012).
\bibitem{lee} P. A. Lee,  arXiv:1401.0519.
\bibitem{agterberg}  D. Agterberg and M. Kashuap, arXiv:1406.4959.
\bibitem{macdonald} W.C. Lee et al,
 Phys. Rev. B 77, 214518 (2008) W.C. Lee and A.H. MacDonald, Phys. Rev. B 78, 174506 (2008).
\bibitem{fomin} I. Fomin, P. Schmitteckert, and P. W\"{o}lfle, Phys. Rev.
Lett. \textbf{69}, 214 (1992).
\bibitem{hao} Z. Hao and A. Chubukov, Phys. Rev. B \textbf{79}, 224513
(2009).
\bibitem{ilya_1} I. Eremin, G. Zwicknagl, P. Thalmeier, and P. Fulde, Phys. Rev. Lett. 101, 187001 (2008).
\bibitem{kontani} S. Onari and H. Kontani, Phys. Rev. B \textbf{84}, 144518
(2011).
\bibitem{peter_book} D. Vollhardt and P. W\"{o}lfle, ``The Superfluid Phases of Helium 3'', Taylor and Francis, London, (1990).
\bibitem{sc} see e.g., I.I. Mazin and J. Schmalian,  Physica C,
\textbf{469}, 614 (2009); S. Graser, T. A. Maier, P. J. Hirshfeld, D. J. Scalapino, New
J. Phys. \textbf{11}, 025016 (2009); K. Kuroki, H. Usui, S. Onari, R. Arita, and H. Aoki, Phys. Rev. B
79, 224511 (2009); J-P Paglione and R.L. Greene, Nature Phys. {\bf 6}, 645 (2010);
D.N. Basov and A.V. Chubukov, Nature Physics {\bf 7}, 241 (2011); P.J. Hirschfeld, M.M. Korshunov, and I.I. Mazin, Reports
on Progress in Physics 74, 124508 (2011); H.H. Wen and S. Li, Annu. Rev. Condens. Matter Phys.,  {\bf 2}, 121 (2011); A.V. Chubukov, Annual Review of Condensed Matter Physics 3, 57 (2012).
 \bibitem{vavilov}  A.V. Chubukov, M.G. Vavilov, and A.B. Vorontsov, Phys. Rev. B 80, 140515(R) (2009);
A. F. Kemper, T. A. Maier, S. Graser, H.-P. Cheng, P. J. Hirschfeld, and D. J. Scalapino
 New J. Phys. 12 073030 (2010).
 \bibitem{raghu} S. Raghu, Xiao-Liang Qi, Chao-Xing Liu, D. Scalapino, and Shou-Cheng Zhang, Phys.Rev.B 77 220503(R) (2008);
 A. V. Chubukov, D. Efremov, and I. Eremin,
Phys. Rev. B \textbf{78}, 134512 (2008).
\bibitem{raman}
M. Khodas, A.V. Chubukov, and G. Blumberg, Phys. Rev. B \textbf{89}, 245134 (2014).
\bibitem{RG}
A. V. Chubukov, D. Efremov, and I. Eremin, Phys. Rev. B \textbf{78}, 134512 (2008).
\end{thebibliography}
\end{document}